\documentclass[11pt]{article}
\usepackage{jheppub}
\usepackage[T1]{fontenc}
\usepackage{amssymb}
\usepackage{mathtools}
\usepackage{stackengine}
\usepackage{scalerel}

\usepackage{hyperref}
\usepackage{epsfig}
\usepackage{amsmath,latexsym,amssymb}
\usepackage{graphicx}
\usepackage{braket}
\usepackage{pdfsync}
\usepackage{framed}
\usepackage{mathtools}

\usepackage{jheppub}
\usepackage[T1]{fontenc}
\usepackage{amssymb}
\usepackage{mathtools}
\usepackage{amsmath}
\usepackage{bm}
\usepackage{stackengine}

\usepackage{scalerel}
\usepackage{eufrak}
\usepackage{framed}

\usepackage{mathtools}

\usepackage{pstricks}

\usepackage{amsmath}
\usepackage{bbold}
\usepackage{bm}
\usepackage{latexsym}
\usepackage{braket}
\usepackage{slashed}
\usepackage{graphicx,booktabs,multirow}
\usepackage{eufrak}
\numberwithin{equation}{section}
\usepackage{color}
\usepackage{pstricks}
\usepackage{amssymb}

\makeatletter
\newcommand{\doublewidetilde}[1]{{%
  \mathpalette\double@widetilde{#1}%
}}
\newcommand{\double@widetilde}[2]{%
  \sbox\z@{$\m@th#1\widetilde{#2}$}%
  \ht\z@=.9\ht\z@
  \widetilde{\box\z@}%
}
\makeatother

\usepackage{hyperref}
\usepackage{slashed}
\usepackage{epsfig}
\usepackage{amsmath,latexsym,amssymb}
\usepackage{graphicx}
\usepackage[latin1]{inputenc}
\usepackage{braket}
\usepackage{pdfsync}
\usepackage{framed}

\usepackage{mathtools}

\def\be{\begin{equation}}
\def\ee{\end{equation}}
\def\ba{\begin{eqnarray}}
\def\ea{\end{eqnarray}}

\newcommand{\bz}{\bar{z}}

\newcommand{\bh}{\bar{h}}

\newcommand{\comment}[1]{}

\newcommand{\eea}{\end{eqnarray}}

\author{
Tomasz R.\ Taylor${}^{1,2}$,\, Bin Zhu${}^{3}$\\[0.5cm]
 $^1${\it Department of Physics,
  Northeastern University, Boston, MA 02115, USA}\\
  $^2${\it Faculty of Physics, University of Warsaw, ul. Pasteura 5, 02-093 Warsaw, Poland}\\
$^3${\it School of Mathematics and Maxwell Institute for Mathematical Sciences,\\ University of Edinburgh,
EH9 3FD, UK }\\[0.2cm]
}

\emailAdd{taylor@neu.edu}
\emailAdd{bzhu@exseed.ed.ac.uk}
\title{\vspace{-2.0cm}\hskip 1cm $\mathbf{w_{1+\infty}}$ Algebra with a Cosmological Constant\\ \hskip 4 cm and the Celestial Sphere}

\abstract{\hfill\\ It is shown that in the presence of a nonvanishing cosmological constant, Stro{\nolinebreak}minger's  infinite-dimensional $\mathrm{w_{1+\infty}}$ algebra of soft graviton  symmetries is modified in a simple way. The deformed algebra contains a subalgebra generating $ SO(1,4)$ or $SO(2,3)$ symmetry groups of $\text{dS}_4$ or $\text{AdS}_4$, depending on the sign of the cosmological constant. The transformation properties of soft gauge symmetry currents under the deformed $\mathrm{w_{1+\infty}}$ are also discussed.

\vskip 4 cm}
\notoc
\makeatletter
\gdef\@fpheader{}
\makeatother

\begin{document}
\maketitle
\noindent \section{}\vskip -1cm
{\it Introduction.} --- The conservation laws reflect the symmetries of nature and provide a key to understanding the physical universe. What was less appreciated until few years ago is the importance of a rather specialized area of quantum field theory and gravity devoted to studying the physical processes involving ``soft'' particles, with very low energies.  The zero energy limit of the scattering amplitudes involving soft gauge bosons, gravitons and other particles are described by ``soft theorems'' \cite{Weinberg:1965nx}. The long wavelengths of soft particles allow probing the large scale structure of universe, particularly the past and future asymptotic infinities. Hence, as shown by Stro{\nolinebreak}minger and collaborators \cite{Strominger:2013lka,Strominger:2013jfa,He:2014laa,Kapec:2014opa,Strominger:2014pwa}, soft theorems are closely related to the conservation laws and symmetries. Actually, every single (known) soft theorem in asymptotically flat spacetime  has been associated with an infinite number of symmetries of the celestial sphere at null infinity. These include Poincar\'e and extended BMS symmetries. \cite{Bondi:1962px,Sachs:1962wk,Barnich:2009se}. The connection between soft theorems and asymptotic symmetries has laid the foundations for the celestial holography program which aims at describing four-dimensional physics in terms of a two-dimensional conformal field theory on the celestial sphere (CCFT) \cite{Raclariu:2021zjz}.

Two tears ago, Strominger performed a systematic study of the symmetries associated with soft particles carrying positive helicities \cite{Strominger:2021lvk}. He showed that all these symmetries are encompassed in an infinite-dimensional $\mathrm{w_{1+\infty}}$ algebra.  $\mathrm{w_{1+\infty}}$ was extracted from the algebra of soft currents encoded in the operator product expansion (OPE) of celestial primary operators associated with gravitons and gauge bosons \cite{Guevara:2021abz}. In CCFT, OPEs can be obtained from the collinear limits of (celestial) scattering amplitudes \cite{Fan:2019emx,Pate:2019lpp}. Strominger's results follow from tree-level amplitudes evaluated in flat spacetime with vanishing cosmological constant.\footnote{Various aspects of soft symmetry algebras are discussed in Refs.\cite{Ball:2021tmb,Himwich:2021dau, Mago:2021wje,Costello:2022wso,Fan:2022vbz,Casali:2022fro,Bu:2022iak,Stieberger:2022zyk,Bittleston:2022jeq,Monteiro:2022xwq,Banerjee:2023rni,Taylor:2023bzj, Garner:2023izn,Bittleston:2023bzp,Melton:2022fsf,Drozdov:2023qoy, Lipstein:2023pih,Costello:2023hmi,Adamo:2023zeh,Ball:2023ukj,Krishna:2023ukw,Banerjee:2023jne,Banerjee:2023bni}}

In Friedman-Lema\^itre cosmology, the observed accelerated expansion of universe can be accounted for  by a positive value of the cosmological constant $\Lambda\approx 10^{-52}$ m$^{-2}$ \cite{pdg}. Hence, universe  is not asymptotically flat -- it is asymptotically de Sitter (dS), at least in the future. In this letter, we construct an algebra similar to Strominger's $\mathrm{w_{1+\infty}}$, ``deformed'' by a nonvanishing cosmological constant. Instead of Poincar\'e, it contains a subalgebra generating the $SO(1,4)$ symmetry group of four-dimensional de Sitter spacetime. The modifications of $\mathrm{w_{1+\infty}}$ are obtained by analyzing OPEs associated with the collinear limits of gravitons and gauge bosons, now corrected by de Sitter curvature.

{\it The idea.} --- The idea originates from the recent work of Alday, Hansen and Silva, who
computed the amplitude for the scattering of four gravitons on $\mathrm{AdS_5\times S_5}$ \cite{Alday:2022uxp,Alday:2022xwz,Alday:2023jdk,Alday:2023mvu}. Although no rigorous definition of the S-matrix exists for non-asymptotically flat spacetimes like AdS,
they formally used AdS/CFT correspondence \cite{Maldacena:1997re} and expanded the amplitude in the inverse curvature radius $R^{-2}\propto\Lambda$. In the limit $\Lambda\to 0$, they obtained the well known Virasoro-Shapiro amplitude. The subleading term is of order ${\cal O}(\Lambda)$. Instead of considering it as a part of a full-fledged AdS S-matrix element, we can consider it as a curvature-induced correction the scattering amplitude in flat spacetime.  For comparison, the proton-proton cross sections measured at the LHC also receive similar corrections although protons do not fly in from the cosmological horizon but come from a bottle of hydrogen stored in Meyrin, Switzerland.  The subleading term has a very interesting property which becomes transparent after taking the string zero slope limit, while keeping fixed the gravitational coupling constant $\kappa$. Then, for the gravitons in the $(-\,-+\,+)$, i.e.\ in the MHV helicity configuration,
\be A^{(1)}(s,t,u)=\kappa^2\Lambda\frac{\langle 12\rangle^{6}[43]}{\langle 13\rangle\langle 14\rangle\langle 23\rangle\langle 24\rangle\langle 34\rangle}\Big(\frac{1}{s}+\frac{1}{t}+\frac{1}{u}\Big),\ee
where $s,t,u$ are the Mandelstam variables and for the momentum spinors, we used the notation of Ref.\cite{Taylor:2017sph}. We are interested in the limit of collinear (light-like) momenta $p_3$ and $p_4$, when $s\to 0$. For our purposes, it is convenient to parametrize momenta in terms of light-cone energies $\omega$ and complex coordinates $z$ on the celestial sphere \cite{Raclariu:2021zjz}. Then, for two arbitrary light-like momenta $p_i$ and $p_j$,
\be
\langle ij\rangle=\sqrt{\omega_i\omega_j}(z_i-z_j)\ ,\quad [ij]=\sqrt{\omega_i\omega_j}(\bz_j-\bz_i)\ , \quad 2p_ip_j=\omega_i\omega_j|z_i-z_j|^2\ .\ee
In order to define the collinear limit, we complexify the momenta, so that $z$ and $\bz$  can be considered as independent complex variables \cite{Taylor:2017sph}. The combined momentum of the collinear pair is defined as $P=p_3+p_4$. In the collinear limit $z_3\to z_4$ (while keeping $\bz_3$ and $\bz_4$ fixed) and $P^2\to 0$. The leading collinear singularity has the form
\be A^{(1)}(s,t,u)\approx \kappa\Lambda\frac{\omega_P^2}{\omega_3^2\omega_4^2(z_3-z_4)^2}\left(\kappa\frac{\langle 12\rangle^{6}}{\langle 1P\rangle^2\langle 2P\rangle^2}\right),\label{dbp}\ee
therefore we obtain the three-graviton amplitude (enclosed in the brackets) times the collinear factor with a double pole $(z_3- z_4)^{-2}$. This is a stronger collinear singularity than the single pole encountered in flat spacetime.

{\it Graviton OPEs.} --- The leading term in the OPE of the primary CCFT operators $G^{+}_{\Delta}(z)$ with dimensions $\Delta$, associated with the (positive helicity) gluons can be extracted from the amplitude (\ref{dbp}) in the same way as in Ref.\cite{Fan:2019emx}. We obtain
\be G^{+}_{\Delta_3}(z_3,\bz_3)\,G^{+}_{\Delta_4}(z_4,\bz_4)\sim \kappa\Lambda \frac{B(\Delta_3-2,\Delta_4-2)}{z_{34}^2}G^{+}_{\Delta_3+\Delta_4-2}(z_4,\bz_4)\ ,\label{ope1}\ee
where $z_{34}=z_3-z_4$. One can also extract a single pole term $\sim z_{34}^{-1}$, which is necessary for the symmetry of the operator product under $3\leftrightarrow 4$.
There is, however, a problem with this OPE. If one proceeds along the lines of Ref.\cite{Guevara:2021abz}, and extracts the algebra of soft currents associated with the graviton operators, it fails Jacobi identity. It is not difficult though, to find a slight modification of the OPE  coefficients leading to consistent double and single pole singularities:
\begin{align}
G^+_{\Delta_3}(z_3,\bar{z}_3)G^+_{\Delta_4}(z_4,\bar{z}_4) = & -\frac{\kappa}{2} \frac{\bar{z}_{34}}{z_{34}} B(\Delta_3-1,\Delta_4-1)  G^+_{\Delta_3+\Delta_4}(z_4,\bar{z}_4)  \nonumber\\[1mm]
&+ \frac{\kappa \, \Lambda}{2} \frac{\Delta_3+\Delta_4}{z_{34}^2} B(\Delta_3-2,\Delta_4-2) G^+_{\Delta_3+\Delta_4-2}(z_4,\bar{z}_4) \nonumber \label{ope2}\\[1mm]
&+\frac{\kappa \, \Lambda}{2} \frac{\Delta_3}{z_{34}} B(\Delta_3-2,\Delta_4-2) \partial G^+_{\Delta_3+\Delta_4-2}(z_4,\bar{z}_4) \, ,
\end{align}
where, for completeness, we also included, in the first term, the  contribution of the zero slope limit of the Virasoro-Shapiro amplitude in flat spacetime. The coefficient of the double pole term,
see the second term on the r.h.s.\ of Eq.(\ref{ope2}),
contains an extra factor of $(\Delta_3+\Delta_4)/2$ as compared to the collinear limit (\ref{ope1}).
We can only speculate that it is due to a modified form of the momentum conservation law in curved spacetime. Indeed, as shown below, it will change the commutation relations of the ``momentum'' operators in a way expected for a spacetime with constant curvature.

{\it From OPEs to cosmological  $\mathrm{w_{1+\infty}}$.} --- After including the antiholomorphic descendants in the OPE of Eq.(\ref{ope2}), it acquires the form:
\begin{align}
G^+_{\Delta_3}(z_3,\bar{z}_3)G^+_{\Delta_4}(z_4,\bar{z}_4) =& -\frac{\kappa}{2} \frac{1}{z_{34}} \sum_{n=0}^{\infty}B(\Delta_3-1+n,\Delta_4-1) \frac{(\bar{z}_{34})^{n+1}}{n!}\bar{\partial}^n G^+_{\Delta_3+\Delta_4}(z_4,\bar{z}_4)  \nonumber\\[1mm]
&+ \frac{\kappa \, \Lambda}{2} \frac{\Delta_3+\Delta_4}{z_{34}^2} \sum_{n=0}^{\infty}B(\Delta_3-2+n,\Delta_4-2) \frac{(\bar{z}_{34})^{n}}{n!}\bar{\partial}^nG^+_{\Delta_3+\Delta_4-2}(z_4,\bar{z}_4) \nonumber\\[1mm]
&+\frac{\kappa \, \Lambda}{2} \frac{\Delta_3}{z_{34}} \sum_{n=0}^{\infty}B(\Delta_3-2+n,\Delta_4-2) \frac{(\bar{z}_{34})^{n}}{n!}\partial\, \bar{\partial}^nG^+_{\Delta_3+\Delta_4-2}(z_4,\bar{z}_4) \, . \label{eq:deformOPE}
\end{align}
Next, we define
the conformally soft graviton operators,
\be H^k=\lim_{\epsilon\to 0}\epsilon G^+_{k+\epsilon}\ , ~~\qquad k=2,1,0,-1,\dots,\label{defh}\ee
with  conformal weights $\{h,\bh\}=\{(k+2)/2,(k-2)/2\}$. We represent them as truncated antiholomorphic series,
\be\label{trunc}
H^k(z,\bar{z}) = \sum_{n=\frac{k-2}{2}}^{\frac{2-k}{2}} \frac{H^{k}_n(z)}{\bar{z}^{n+\frac{k-2}{2}}}\ ,
\ee
and further expand the holomorphic coefficients: \be H^{k}_n(z)
= \sum_{a=-\infty}^\infty \frac{H^k_{a,n}}{z^{a+\frac{k+2}{2}}}
\ee
The OPE of Eq.(\ref{eq:deformOPE}) translates into the following algebra of soft currents:
\begin{align}
[H^k_{a,m},H^l_{b,n}] =& -\frac{\kappa}{2}[n(2-k)-m(2-l)] \frac{\left( \frac{2-k}{2} -m+ \frac{2-l}{2}-n-1\right)!\left( \frac{2-k}{2} +m+ \frac{2-l}{2}+n-1\right)! }{\left( \frac{2-k}{2}-m\right)!\left( \frac{2-l}{2}-n\right)!\left( \frac{2-k}{2}+m\right)!\left( \frac{2-l}{2}+n\right)!} H^{k+l}_{a+b, m+n}  \nonumber\\[2mm]
&+\frac{\kappa \, \Lambda}{2}\left(l\, a -k \, b\right) \frac{\left( \frac{2-k}{2} -m+ \frac{2-l}{2}-n\right)!\left( \frac{2-k}{2} +m+ \frac{2-l}{2}+n\right)! }{\left( \frac{2-k}{2}-m\right)!\left( \frac{2-l}{2}-n\right)!\left( \frac{2-k}{2}+m\right)!\left( \frac{2-l}{2}+n\right)!} H^{k+l-2}_{a+b, m+n}   \, . \label{eq:deformH}
\end{align}
To make connection with Strominger's $\mathrm{w_{1+\infty}}$, we define
\be
w^{p}_{a,m} = \frac{1}{\kappa} (p-m-1)!(p+m-1)! H^{-2p+4}_{a,m} \, , \label{eq:wfromH}\ee
where $p$  run over the positive half integers
\be
p= 1, \, \frac{3}{2}, \, 2, \, \frac{5}{2} \, \dots \ ,\label{eq:prange}
\ee
and the condition of the truncated antiholomorphic mode expansion (\ref{trunc}) turns into the following constraint on the indices labelled by $m$:
\be
1-p\leq m\leq p-1 \, . \label{eq:mrange}
\ee
The indices $a$, associated with the holomorphic modes, are integer for integer $p$  and half-integer for half-integer $p$, similar to $m$, but their range  is not restricted:\be a=-\infty,\dots,p-1,p,p+1,\dots,\infty .\label{arange}\ee
The algebra (\ref{eq:deformH}), written in terms of $w$ generators, becomes
\be
 \, \, [w^{p}_{a,m} , \, w^q_{b,n}] = [m(q-1)-n(p-1)] w^{p+q-2}_{a+b,m+n} - \Lambda [a(q-2)-b(p-2)] w^{p+q-1}_{a+b,m+n} \, . \label{eq:deform1fromOPE}
\ee
It is easy to check that it satisfies Jacobi identity and closes within the range of indices given in Eqs.(\ref{eq:prange}-\ref{arange}).

{\it Properties of deformed algebra}. --- In order to understand the structure of the deformed $\mathrm{w_{1+\infty}}$ algebra (\ref{eq:deform1fromOPE}) and the role of the cosmological constant, we note that it contains a closed subalgebra of 10 generators: $w^1_{a,0}$ with $a=-1,0,1$, $w^{\frac{3}{2}}_{\pm\frac{1}{2},\pm\frac{1}{2}}$, and $w^2_{0,m}$ with $m=-1,0,1$. It reads:
\begin{align}&[w^1_{a,0},w^2_{0,m}] = 0\ ,\\[1mm]
&[w^1_{a,0}, w^1_{b,0}] = \Lambda (a-b)w^1_{a+b,0}\ , \\[1mm]
&[w^2_{0,m},w^2_{0,n}] = (m-n)w^2_{0,m+n} \ ,\\
&[w^1_{a,0},w^{\frac{3}{2}}_{k,l} ]= \Lambda\left(\frac{a}{2}-
k\right) w^{\frac{3}{2}}_{a+k,l} \ ,\\ &
[w^2_{0,m},w^{\frac{3}{2}}_{k,l} ]= \left( \frac{m}{2}-l\right) w^{\frac{3}{2}}_{k,m+l} \ , \\
&[w^{\frac{3}{2}}_{i,j}, w^{\frac{3}{2}}_{k,l}] = \frac{1}{2}(j-l) w^{1}_{i+k,j+l}+\frac{\Lambda}{2}(i-k) w^2_{i+k,j+l}\ . ~~~~~~~~~~~~~~~~~~~~~~~\end{align}
In Strominger's $\mathrm{w_{1+\infty}}$,  $w^1_{a,0}$ are c-number operators. In the present algebra, however, they do not commute. To see it in a more transparent way, we define
\be
w^1_{a,0} = \Lambda \, L_a \, , \quad w^2_{0,m} = \bar{L}_m\, ,\quad
w^{\frac{3}{2}}_{k,l} = P_{k,l} \, . \label{eq:idPkl}
\ee
In terms of these operators, the above algebra reads
\begin{align}&
[L_a,\bar{L}_m] = 0 \ ,  \label{eq:subal3}\\[1mm] &
[L_a,L_b] = (a-b)L_{a+b} \, , \label{eq:subal1}\\[1mm] &
[\bar{L}_m,\bar{L}_n]  = (m-n) \bar{L}_{m+n} \, ,\label{eq:subal2}\\[1mm]  &
[L_a,P_{k,l}] = \left(\frac{a}{2}-k\right) P_{a+k,l} \, ,\label{eq:subal4}\\ &
[\bar{L}_{m}, P_{k,l}] = \left(\frac{m}{2}-l\right) P_{k,m+l} \, ,\label{eq:subal5} \\ &
[P_{i,j},P_{k,l}] = \Lambda \, j \, \delta_{j,-l} \, L_{i+k} +\Lambda \, i\, \delta_{i,-k} \, \bar{L}_{j+l} \  .~~~~~~~~~~~~~~~~~~~~~~~~~~~~~~~~~\label{eq:subal6}
\end{align}
In the limit of $\Lambda=0$, it is the Poincar\'e subalgebra of extended BMS symmetry,
with four translations $P_\mu\, ,\mu=0,1,2,3$, defined by,
\begin{align}
P_{-\frac{1}{2},-\frac{1}{2}} &= P_0+P_3 \, , \quad P_{-\frac{1}{2},\frac{1}{2}} = P_1-iP_2 \, , \nonumber\\
P_{\frac{1}{2},-\frac{1}{2}} &= P_1+iP_2 \, , \quad P_{\frac{1}{2},\frac{1}{2}} = P_0-P_3 \, ,
\label{eq:Pmodes}
\end{align}
and with six Virasoro operators $L_{1,0,1}\, , \bar L_{-1,0,1} $ related to the Lorentz generators $M_{\mu\nu}=-M_{\nu\mu}$ in the following way \cite{Stieberger:2018onx,Fotopoulos:2019vac}:
\begin{align}
M_{23}+iM_{10} &= -L_{-1}+L_1 \, , \qquad  -M_{23}+iM_{10} = -\bar{L}_{-1} +\bar{L}_1 \, ,  \\
M_{20} +iM_{13} &= -L_{-1}-L_1 \, , \qquad -M_{20}+iM_{13} = -\bar{L}_{-1} -\bar{L}_1 \, , \\
M_{21}+iM_{03} &= -2 L_0 \, ,~~~\qquad \quad -M_{21}+iM_{03} = -2\bar{L}_0 \, ,
\end{align}
In the presence of a nonvanishing cosmological constant, however, translations do not commute and the Poincar\'e algebra is deformed,  depending on the sign of $\Lambda$,  to the algebra generating $SO(1,4)$ or $SO(2,3)$ symmetry groups of dS$_4$ or AdS$_4$, respectively. To see this, we define
\be P_\mu = \frac{\sqrt{|\Lambda|}}{2} M_{\mu 4} \, . \label{eq:PtoMi4}
\ee
Then the algebra of Eqs.(\ref{eq:subal3}-\ref{eq:subal6}) can be written as
\be
[M_{\mu\nu},M_{\rho\lambda}] = i(\eta_{\mu\rho}\,M_{\nu\lambda} + \eta_{\nu\lambda} M_{\mu\rho} -\eta_{\nu\rho} M_{\mu\lambda} -\eta_{\mu\lambda} M_{\nu\rho}) \, ,
\ee
with
\be\eta_{\mu\nu}=\mathit{diag}\big(1,-1,-1,-1,-\mathit{sign}(\Lambda)\big).\ee
The cosmological $\mathrm{w_{1+\infty}}$ algebra of Eq.(\ref{eq:deform1fromOPE}) provides an infinite-dimensional extension of dS$_4$ and AdS$_4$ symmetries.

{\it Gauge theory}. --- We started this discussion from the collinear limit of curvature corrections to the graviton scattering amplitudes \cite{Alday:2022uxp,Alday:2022xwz,Alday:2023jdk,Alday:2023mvu}. It would be very interesting to study gauge theories coupled to gravity in a similar way, in particular curvature corrections to graviton-gauge boson interactions in Einstein-Yang-Mills theory. Here, we also expect a double pole in the OPE of the positive helicity graviton operator $G^+_{\Delta}(z,\bz)$ and the gauge boson operator $O^{+d}_{\Delta}(z,\bar{z})$, where $d$ labels the group index.  It would modify the corresponding OPE in the following way:
\begin{align}
G^+_{\Delta_1}(z_1,\bar{z}_1) O^{+d}_{\Delta_2}(z_2,\bar{z}_2) =& -\frac{\kappa}{2} \frac{\bar{z}_{12}}{z_{12}} B(\Delta_1-1,\Delta_2) O^{+d}_{\Delta_1+\Delta_2}(z_2,\bar{z}_2) \nonumber\\[1mm]
&+ \frac{\kappa \Lambda}{2} \, \frac{\Delta_1+\Delta_2-1}{z_{12}^2} \, B(\Delta_1-2,\Delta_2-1) O^{+d}_{\Delta_1+\Delta_2-2}(z_2,\bar{z}_2) \nonumber\\
&+\frac{\kappa \Lambda}{2}\, \frac{\Delta_1}{z_{12}} \, B(\Delta_1-2,\Delta_2-1) \partial O^{+d}_{\Delta_1+\Delta_2-2}(z_2,\bar{z}_2)  \, .\label{ope5}
\end{align}
To see what is the corresponding deformation of the symmetry algebra, we define
the conformally soft gluon operators,
\be
R^{k,d} = \lim_{\epsilon\rightarrow0} \epsilon \, O^{+d}_{k+\epsilon}, \quad k=1,0,-1, \dots \,
\ee
with conformal weights $\{h,\bh\}=\{(k+1)/2,(k-1)/2\}$. We represent them as a truncated holomorphic series:
\be
R^{k,d}(z,\bar{z}) = \sum_{n=\frac{k-1}{2}}^{\frac{1-k}{2}} \frac{R^{k,d}_n(z)}{\bar{z}^{n+\frac{k-1}{2}}}\ ,\ee 
and further expand the holomorphic coefficients,
\be R^{k,d}_n(z)= \sum_{a=-\infty}^{\infty} \frac{R^{k,d}_{a,n}}{z^{a+\frac{k+1}{2}}} \, .
\ee
The algebra of the soft currents following from the OPE (\ref{ope5}) has the form:
\begin{align}
[H^k_{a,m},R^{l,d}_{b,n}] =& -\frac{\kappa}{2}[n(2-k)-m(1-l)] \frac{\left( \frac{2-k}{2} -m+ \frac{1-l}{2}-n-1\right)!\left( \frac{2-k}{2} +m+ \frac{1-l}{2}+n-1\right)! }{\left( \frac{2-k}{2}-m\right)!\left( \frac{1-l}{2}-n\right)!\left( \frac{2-k}{2}+m\right)!\left( \frac{1-l}{2}+n\right)!} R^{k+l,d}_{a+b, m+n}  \nonumber\\[1mm]
&+\frac{\kappa \, \Lambda}{2}\left((l-1)\, a -k \, b\right) \frac{\left( \frac{2-k}{2} -m+ \frac{1-l}{2}-n\right)!\left( \frac{2-k}{2} +m+ \frac{1-l}{2}+n\right)! }{\left( \frac{2-k}{2}-m\right)!\left( \frac{1-l}{2}-n\right)!\left( \frac{2-k}{2}+m\right)!\left( \frac{1-l}{2}+n\right)!} R^{k+l-2,d}_{a+b, m+n}   \, . \label{eq:deformHR}
\end{align}
Upon the redefinition written in Eq.(\ref{eq:wfromH}) and
\be
S^{q,d}_{a,m} = (q-m-1)!(q+m-1)!R^{3-2q,d}_{a,m} \, ,
\ee
the commutators (\ref{eq:deformHR}) become
\be
\left[ w_{a,m}^p, S^{q,d}_{b,n}\right] = [m(q-1)-n(p-1)] S^{p+q-2,d}_{a+b,m+n} - \Lambda [ a(q-1)-b(p-2)] S^{p+q-1,d}_{a+b,m+n} \, . \label{eq:deformwS}
\ee

While gravitational interactions are affected by curvature, we do not expect corrections to pure Yang-Mills thery, therefore the $S$-algebra of soft gauge currents should remain in its original form
\cite{Strominger:2021lvk},
\be
\left[S^{p,d}_{a,m}, S^{q,e}_{b,n}\right] = -i f^{deg} S^{p+q-1,g}_{a+b,m+n} \, . \label{eq:SS}
\ee
The soft currents $S^{1,d}_{0,0}$ satisfy
\be
\left[S^{1,d}_{0,0},S^{1,e}_{0,0}\right] = -if^{deg}S^{1,g}_{0,0} \, .
\ee
and generate global gauge transformations.
From Eq.(\ref{eq:deformwS}), it follows that
\be
\left[ L_a, S^{1,d}_{0,0}\right] = \left[\bar{L}_m, S^{1,d}_{0,0}\right] = \left[P_{k,l}, S^{1,d}_{0,0} \right]= 0 \, .
\ee
Although we do not have a solid argument supporting the cosmological deformation written in Eq.(\ref{eq:deformwS}), it is easy to check that the full symmetry algebra of Einstein-Yang-Mills systems, written in Eqs.(\ref{eq:deform1fromOPE}), (\ref{eq:deformwS}) and (\ref{eq:SS}), satisfies all Jacobi identities.

{\it Discussion.} --- In this letter, we constructed a cosmological deformation of $\mathrm{w_{1+\infty}}$ algebra, by including commutator terms proportional to the cosmological constant. As a result, the Poincar\'e subalgebra was replaced by the symmetry algebra of dS or AdS, depending on the sign of the cosmological constant. It is striking that this deformation was extracted from the singularity structure of the graviton scattering amplitudes in spacetime with constant curvature, although a slight modification of the corresponding graviton OPEs was necessary to ensure a self-consistent algebra. It would be very interesting to uncover a deeper reason for this modification. A precise connection to the construction
of Alday, Hansen and Silva \cite{Alday:2022uxp,Alday:2022xwz,Alday:2023jdk,Alday:2023mvu} remains to be understood. In particular, their higher-order curvature corrections contain  higher order poles in Mandelstam variables. The physical interpretation of these poles is not clear.

It would be also interesting to see how the cosmological $w_{1+\infty}$ algebra is related to the asymptotic symmetries of dS spacetime, discussed in Refs.\cite{Strominger:2001pn,Anninos:2010zf,Anninos:2011jp,Compere:2019bua,Compere:2020lrt,Compere:2023ktn,Mao:2019ahc,Bonga:2023eml}. Furthermore,
one might utilize the methods developed in \cite{Freidel:2021ytz} to find a realization of the cosmological $w_{1+\infty}$ algebra on the gravitational phase space.

\section*{Acknowledgements}
This work was inspired by a talk given by John Schwarz at the kickoff workshop of Simons  Collaboration on Celestial Holography  (Cambridge, MA, October 26-29, 2023), in which he proposed a modification of the OPEs similar to Eq.(\ref{ope2}). We would like to thank John Schwarz and Andy Strominger for comments on the manuscript and to Fernando Alday for useful correspondence.
TRT is supported by the National Science Foundation
under Grants Number PHY-1913328 and PHY-2209903, by the
NAWA Grant
``Celestial Holography of Fundamental Interactions'' and
by the Simons  Collaboration on Celestial Holography under grant MP-SCMPS-00001550-05.
Any opinions, findings, and conclusions or
recommendations expressed in this material are those of the authors and do not necessarily
reflect the views of the National Science Foundation. BZ is supported by the Royal Society.
\end{document}